\documentclass[%
 reprint,  amsmath,amssymb,
 pra, apa]{revtex4-1}

\usepackage{gatez}

\usepackage{graphicx}
\usepackage{dcolumn}
\usepackage{bm}

\usepackage{fancyhdr}
\usepackage{algpseudocode}
\usepackage{algorithm}

\usepackage{comment}

\usepackage{parskip}
\usepackage{tensor}
\usepackage{amsmath}
\usepackage{amssymb}
\usepackage{amsthm}
\usepackage{bbm}
\usepackage{slashed}

\begin{document}


\title{Quantum Computation with Machine-Learning-Controlled Quantum Stuff}

\author{Lucien Hardy}
\author{Adam G.\ M.\ Lewis}
 \email{alewis@perimeterinstitute.ca}
\affiliation{Perimeter Institute for Theoretical Physics, \\ 31 Caroline Street North, Waterloo, Ontario N2L 2Y5, Canada}

\date{\today}

\begin{abstract}
We describe how one may go about performing quantum computation with arbitrary \lq\lq quantum stuff", as long as it has some basic physical properties. Imagine a long strip of stuff, equipped with regularly spaced wires to provide input settings and to read off outcomes. After showing how the corresponding map from settings to outcomes can be construed as a quantum circuit, we provide a machine learning algorithm to tomographically ``learn" which settings implement the members of a universal gate set. At optimum, arbitrary quantum gates, and thus arbitrary quantum programs, can be implemented using the stuff.
\end{abstract}

\maketitle


\section{\label{sec:introduction}Introduction}

Imagine we have some \lq\lq stuff" with quantum properties.  Can we use it as a quantum computer? Join us in picturing a long strip of Plasticine-like stuff, whose unknown properties are accessible only via regularly spaced setting and outcome wires.  By appropriate choices of settings, followed by correct interpretations of outcomes, is it possible to implement quantum computation?

If the stuff secretly contains matter known to be capable of quantum computation - say, a row of trapped ions - it is obviously possible. Suppose, for example, that the setting wires were to manipulate the local magnetic field and the timing of measurements, whose results were transmitted by the outcome wires. Any given quantum program could then be implemented, given knowledge of the details.

Normally, one would first decide upon a correspondence between physical and logical operations, and then engineer a computer to respect it. Here, we discuss the reverse task, that of mapping arbitrary quantum computation onto the fixed physics of initially uncharacterized matter: stuff. 

Our iterative approach to this problem begins with a hypothesized map between physical and logical operations. For example, we might guess that feeding first $2$ and then $3$ to the first input wire implements the quantum CNOT gate. This guess is likely to be wrong, but we will demonstrate that given a few assumptions about the internal dynamics of the stuff, it is possible to determine just \emph{how} wrong. We can then converge towards a correct guess by gradient descent, applied to a specially constructed set of neural networks.

The essential problem we are concerned with, the determination of the unknown quantum dynamics of a black box system (in our case, the arbitrary length strip of stuff), might be viewed as a slight generalization of so-called ``quantum process tomography" \cite{Tomography1, Tomography2, Tomography3, Tomography4, Tomography5, Tomography6}. We are aware of two major differences. First, we assume no ability to interact with the quantum stuff except via the classical settings and outcomes on the wires. In particular, we do not presume advance knowledge of how to prepare or to measure states. This echoes previous work on randomized gate benchmarking \cite{knill2008randomized, emerson2007symmetrized, emerson2005scalable, magesan2011scalable}, self-consistent quantum process tomography \cite{SelfConsistentQPT1, SelfConsistentQPT2}, and especially operational process tomography \cite{diMatteoOST}.

Second, we will never explicitly obtain the process matrix of the stuff, but only the ability to map chosen quantum circuits onto it. Presumably, the process matrix would need to be somehow encoded in the weights of any successfully optimized neural networks, but we make no attempt to characterize such an encoding. This situates our work within the rapidly developing field of ``quantum machine learning" \cite{biamonte2017quantum, schuld2015introduction}, and, more specifically, within the subfield concerned with applying machine learning approaches to quantum tomography \cite{MLTom1, MLTom2,  MLTom3, MLTom4, MLTom5, MLTom6}.

In Section \ref{sec:Overview} to follow, we introduce the essential ideas required to construe operations upon quantum stuff as a user-defined circuit mapped onto spacetime. These ideas are then refined and formulated precisely in Section \ref{sec:Circuits}. The iterative process of determining which operations correspond to which circuit, or ``bootstrap tomography", is formulated in Section \ref{sec:bootstrap}. Finally, in Section \ref{sec:MachineLearning}, we propose a machine learning algorithm to realize this tomographic process.

\begin{figure*}[!ht]
\begin{tikzpicture}[scale=0.3]
\stuff[1]{20}{0,0}
\spacetimepoints[1]{19}{19}{0,-25}
\draw[thin,->] (-1.5,-20)--(-1.5,-10) node[midway,left]{$t$};\draw[thin,->](5,-26.5)--(14,-26.5) node[midway, below]{$x$};
%
\begin{scope}[xshift=25cm]
\stuff[1]{19}{0,0}
\spacetimepoints[1]{19}{19}{0,-25}
\clip (0,-25) rectangle (19,-6);
\begin{scope}[xshift=-7.5cm,yshift=-32.5cm]
\octagons{30,36}
\end{scope}
\end{scope}
\draw[thick] ($ (current bounding box.south west)+(-2,-2)$)  rectangle
($(current bounding box.north east)+(2,2)$);
\draw[thick] ($(current bounding box.south)+(1.3,0)$) --
($(current bounding box.north)+(1.3,0)$);
\end{tikzpicture}
\caption{On the left we see a section of a length of stuff with input and output wires placed at regular intervals.  A time-gated sequence of inputs is fed into the input wires and a similar time-gated sequence of outputs is read off the wires.   We represent the input/output at position $x$ and time $t$ by a dot as shown.  On the right we see the same figure with the dots divided up into octagons and squares.  Points lying on the boundary between octagons are assigned to the upper octagon.}
\label{fig:stuffdotsoctagons}
\end{figure*}
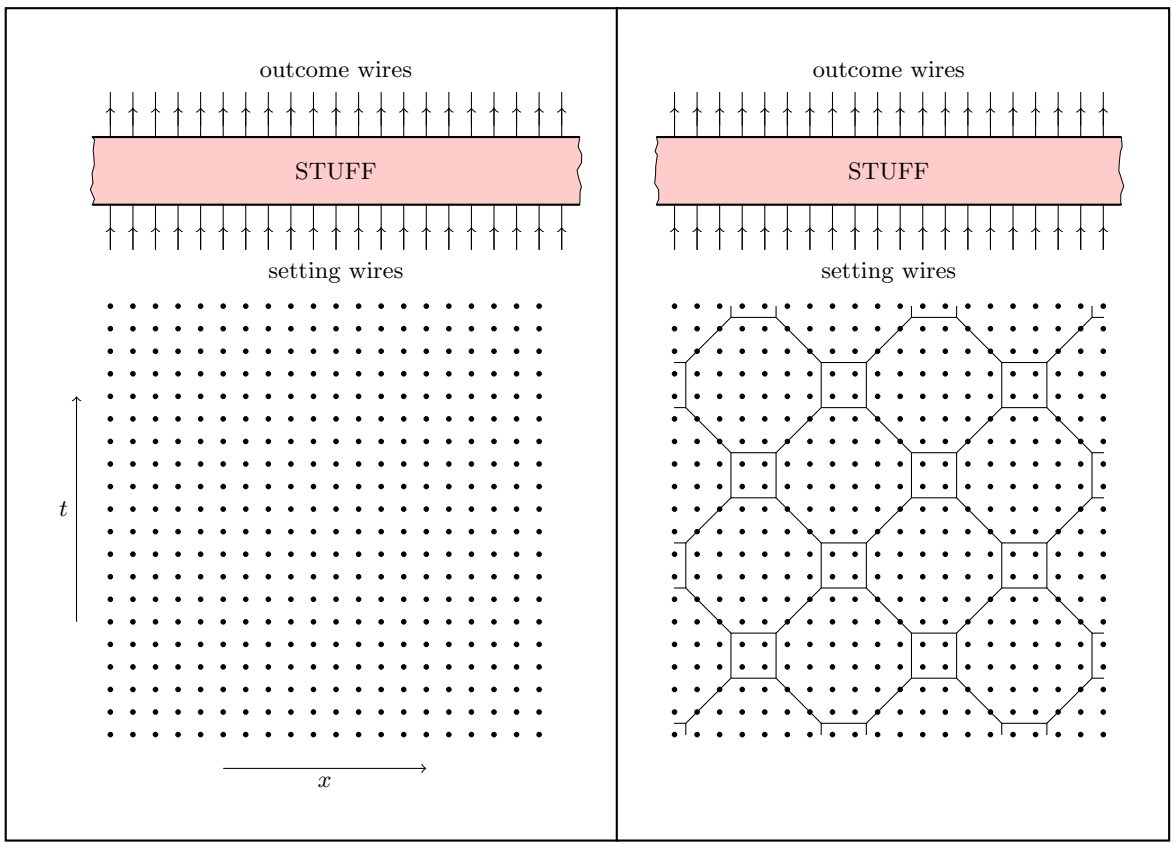

\label{sec:Overview}
\section{Quantum Stuff and computation}
Picture a strip length-$L$ strip of stuff. Fix two wires at each site $x_l=l\Delta X$, $l = (0, 1, 2, \ldots)$, one to send in classical settings, and another to read off classical outcomes. Binning the duration of these interactions into discrete time intervals centred at times $t_n= n\Delta T$, $n = (0, 1, 2, \ldots)$ defines a discrete set of spacetime points, $\mathbf{x}=(x, t)=(l\Delta x, n\Delta t)$, each labelling a setting-outcome pair. We refer this structure of setting-outcome pairs at discrete events as the \emph{computational lattice}.

The physical dynamics of the stuff might be viewed as a map from sequences of settings to probability distributions over sequences of outcomes. To use the stuff as a quantum computer, we need to know the pertinent aspects of that mapping, construed as one between known mathematical objects. This knowledge can be usefully decomposed into that of two functions: an \emph{encoder} to translate a given program - the ``logical input" - into physical settings, and a \emph{decoder} to translate probability distributions over physical outcomes into mathematical objects, the ``logical output". The encoder/decoder pair has functioned correctly if and only if the emitted logical output is, accounting for the probabilistic nature of the physical outcomes, indeed that dictated by the logical input. 

Suppose, for example, we wish to perform the computation $q_\mathrm{out} = 2 q_\mathrm{in}$. Then the encoder must map from the logical input ``$2 \times q_\mathrm{in}$" to some sequence of physical settings. The decoder must then map the corresponding sequence of physical outcomes into the logical output $q_\mathrm{out}$. The encoder/decoder pair is correct, in this case, when the emitted $q_\mathrm{out}$ approaches $2 q_\mathrm{in}$, having smoothed (e.g. averaged over) probabilistic fluctuations.

\begin{table*}[t]
\begin{tabular}{|l|c|}
  \hline
Subset 1 \rule{0pt}{8.1 ex}  &  ~~\Gate{\cnot}  ~ \Gate{\HIgate} ~ \Gate{\IHgate}~ \Gate{\PIgate} ~ \Gate{\IPgate} ~ \Gate{\RIgate} ~ \Gate{\IRgate}~~ \\
  \hline
Subset 2   \rule{0pt}{8.1 ex} & \Gate{\IIgate} ~\SwapGate \\
 \hline
Subset 3\rule{0pt}{8.7 ex}   & \prepZZGate ~ \Gate{\zeroI} ~ \Gate{\oneI}~ \Gate{\Izero} ~ \Gate{\Ione}  \\
  \hline
\end{tabular}
\caption{A set of gates that is universal with respect to the given lattice. $H$ stands for the Hadamard gate, $P$ stands for the phase gate which acts as $\left( \begin{array}{cc} 1 & 0 \\ 0 & i \end{array} \right)$, and $R$ stands for the $\frac{\pi}{8}$ gate on the left which acts as $\left( \begin{array}{cc} 1 & 0 \\ 0 & e^{\frac{\pi}{8}} \end{array} \right)$. }\label{UGStable}
\end{table*}

One might reasonably expect the discovery of a correct encoder/decoder pair to be a fairly daunting task. However, in this case and, as we will see, in general, checking whether a given encoder/decoder pair is correct is quite simple. Furthermore, one can easily construct a smooth measure, or \emph{loss function}, of how far from correct a given encoder/decoder pair is: the RMS error between $q_\mathrm{out}$ and $2q_\mathrm{in}$, for example, or some monotonic function of it \footnote{The loss function will, strictly speaking, be a function\emph{al} of the encoder/decoder pair. These will eventually be represented as neural networks, however, parameterized a finite set of weights. The loss will be a plain old function of these.}. A correct encoder/decoder pair will minimize this loss function. Later, we will detail a means to accomplish this functional minimization automatically on a classical computer, via a bespoke neural network parameterization of the encoder and decoder.

\section{Gates, circuits, and tesselations}
We would like to implement arbitrary quantum computations, not just multiplication by 2. It is well known that arbitrary quantum computations can be decomposed into elemental quantum gates forming a ``universal gate set" (UGS). It would thus be sufficient for our encoder/decoder pair to correctly implement all members of some UGS. 

Quantum gates, however, map between quantum states, not classical information. By assumption, we do not have direct access to the internal quantum dynamics of the stuff. We will thus instead concern ourselves with quantum \emph{circuits}, wirings together of quantum gates that are entirely characterized by classical settings and outcomes. 

\begin{figure*}
\begin{tikzpicture}[scale=0.3]
\begin{scope}[xshift=25cm, yshift=-23.5cm]
\clip (0,-25) rectangle (19,-6);
\begin{scope}[xshift=-7.5cm,yshift=-32.5cm]
\octagons{30,36}
\end{scope} 
\begin{scope}[scale=0.6666] 
\yellowgate[0.2]{\prepZZ}{A}{11.25,-32}  \yellowgate[0.2]{\prepZZ}{B}{23.25,-32}
\yellowgate{\IHgate}{C}{5.25,-26.25} \yellowgate{\cnot}{D}{17.25,-26.25}
\yellowgate{\cnot}{E}{11.25,-20.25} \yellowgate{\zeroI}{H}{23.25,-20.25}
\yellowgate{\Izero}{F}{5.25, -14.25}
\yellowgate{\oneI}{G}{17.25,-14.25}
\RDwireD{A}\LDwireD{A} \RDwireD{B}\LDwireD{B}
\LDwire{C} \lwire{A}{C} \rwire{A}{D} \lwire{B}{D} \RUwire{B}
\LUwire{C}  \rwire{C}{E} \lwire{D}{E} \rwire{D}{H}\RUwire{H}
\LDwire{F} \lwire{E}{F} \rwire{E}{G}  \lwire{H}{G}\RDwire{H}
\RUwire{F}\LUwire{F} \RUwire{G}\LUwire{G}
\end{scope}
\end{scope}
\begin{scope} [yshift=-23.5cm, scale=0.6666]
\yellowgate[0.2]{\prepZZ}{A}{11.25,-32}  \yellowgate[0.2]{\prepZZ}{B}{23.25,-32}
\yellowgate{\IHgate}{C}{5.25,-26.25} \yellowgate{\cnot}{D}{17.25,-26.25}
\yellowgate{\cnot}{E}{11.25,-20.25} \yellowgate{\zeroI}{H}{23.25,-20.25}
\yellowgate{\Izero}{F}{5.25, -14.25}
\yellowgate{\oneI}{G}{17.25,-14.25}
\RDwireD{A}\LDwireD{A} \RDwireD{B}\LDwireD{B}
\LDwire{C} \lwire{A}{C} \rwire{A}{D} \lwire{B}{D} \RUwire{B}
\LUwire{C}  \rwire{C}{E} \lwire{D}{E} \rwire{D}{H}\RUwire{H}
\LDwire{F} \lwire{E}{F} \rwire{E}{G}  \lwire{H}{G}\RDwire{H}
\RUwire{F}\LUwire{F} \RUwire{G}\LUwire{G}
\end{scope}
\draw[thick] ($ (current bounding box.south west)+(-2,-2)$)  rectangle
($(current bounding box.north east)+(2,2)$);
\draw[thick] ($(current bounding box.south)+(-0.7,0)$) --
($(current bounding box.north)+(-0.7,0)$);
\end{tikzpicture}
  \caption{On the left we see an example of a circuit built with gates from our universal gate set.  Note that the circuit is closed off from external influences because, at the bottom, input signals are absorbed by the identity measurement and, at the sides, quantum information coming into the circuit is shunted back out.  On the right we see how gates can be assigned to octagons.}\label{circuitsoctagons}
\end{figure*}
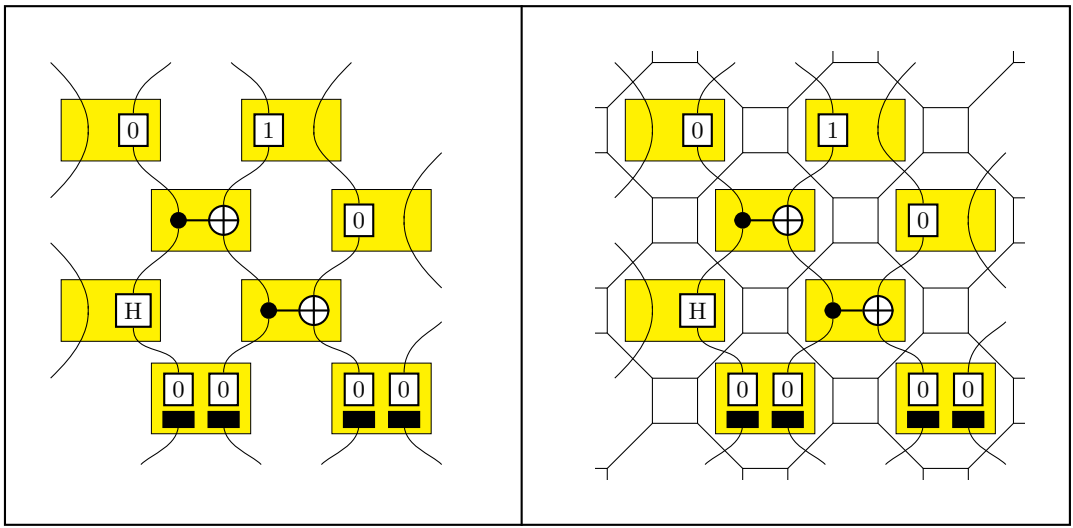

Any given quantum computation may be expressed as a quantum circuit. Doing so further refines the formulation of our task. The logical input thus becomes the classical settings and gate labels defining a particular quantum circuit. The logical output is that circuit's theoretical outcome distribution. The encoder/decoder pair has functioned correctly if the emitted outcome distribution matches the theoretical one. The loss function can be any of various distance metrics between those distributions.

Later we will demonstrate that, given a few physical assumptions about the stuff, there exist ``tomographically complete" sets of quantum circuits. If an encoder/decoder pair, acting upon stuff with the assumed properties, correctly implements all the circuits in a tomographically complete circuit set, it correctly implements all the gates in a UGS. The combination of encoder, stuff, and decoder in that case forms a universal computer. The degree by which it fails to do so can be taken, most primitively, as the summed losses of all the circuits in the set.

We want the stuff not only to implement an arbitrary computation, but to restrict the required number of setting/outcome retrievals to do so. To achieve this, we will require the encoder's implementation of each individual gate to occupy only a finite volume of spacetime. Specifically, we group together events in the computational lattice $\mathbf{x} = \left(l \Delta x, x \Delta t\right)$, depicted in the left panel of Figure \ref{fig:stuffdotsoctagons}, into a \emph{tesselation} of octagons and squares, depicted in the right panel of Figure \ref{fig:stuffdotsoctagons}. 

Each octagon, or \emph{tessel}, will be used to implement one of a universal set of two qubit gates. The qubits for these gates will, under this implementation, be input at the lower slanted edges, and output at the upper slanted edges.  Octagons are convenient because they adjoin to form causal-diamond-like structures. The squares appear by geometric necessity and will implement a fixed ``do nothing" gate, different from the identity. 
We will eventually view the encoder and decoder as neural networks to be trained by a machine learning algorithm. One failure mode of this algorithm will occur if the tesselation encloses too few events per gate to implement the UGS properly. In that case, we can uniformly scale up the tesselation and try again.



We label each octagon by its midpoint, $\mathbf{x}$.  Let $X_\text{octagons}$ be the set of octagon positions we consider in some given tessellation over the length $L$ and some time duration, $T$.

The two qubit gates situated on the octagons will form a regular lattice which we will call the \emph{gate lattice}.  A universal gate set (UGS) must include enough gates to do universal quantum computation \emph{with respect to the gate lattice}.  An example of a universal set of gates, complete with respect to this lattice is given in Table \ref{UGStable}.
Each of the gates in the table is a two-qubit gate.  The first subset of gates are the usual gates included in a UGS adapted for the lattice.  

Because we assume control over the stuff only through the classical settings, we need to include two more subsets of gates not usually mentioned as part of a UGS.
The second subset are the identity and swap gates which allow us to transport qubits around the circuit.  The third subset of gates are the preparation and measurement gates. These gates are non-unitary.  The first gate in subset 3 performs the identity measurement on the incoming qubits and then prepares two qubits, each in the $0$ state.  The second gate in subset 3 projects the left qubit onto the $0$ basis while leaving the right qubit unchanged. We have similar notation for the other gates.  Unlike the other gates in the table, these measurement gates have outcomes associated with them.  

It is necessary to include the second and third subsets of gates in our UGS because we need to train our stuff to implement them the same as any other gate. They do not come for free.

We can connect gates together to form what we will call a \emph{fragment} (denoted by $\mathcal{F}$).  Whenever we build a fragment, there must be some gates having one or two open inputs. We will call such gates \emph{initial gates}.   A circuit, $\mathcal{C}$, is a special case of a fragment for which all initial gates are of the type that ignore any open inputs. The gates in the table provide two ways to do this.  First, the preparation gate in subset 3 simply absorbs incoming quantum systems so they do not affect probabilities for the circuit.  Second, any gates that include an identity map can be used to shunt quantum information coming through an open input back out through an open output so probabilities for the circuit are not affected.  The circuit in Figure \ref{circuitsoctagons} contains examples of both types.  We can calculate a probability for a circuit using the rules of Quantum Theory.  A fragment that is not a circuit is subject to outside influences and so will not necessarily have a probability associated with it.

   We will choose some particular UGS, call it $G$, to proceed.  It does not have to be the one described here, and there might UGS's that are more suited to this project.  However, the gate set must have elements that enable us to close off in the manner just described. There must also be some gates with outcomes, so we can read off the results of the computation.

\section{Building Circuits}
\label{sec:Circuits}
In this section, we describe how sequences of operations upon the stuff may be arranged in such a way to permit comparison with a given quantum circuit.

Thus, within each octagon of the tesselation, we will \emph{encode} (by inputting an appropriate sequence of signals into the setting wires attached to the stuff) and \emph{decode} (by selecting on an appropriate sequence of output signals from the outcome wires attached to the stuff) in an attempt to implement a putative element of the UGS, $G$. Initially we do not know what the appropriate encoding and decoding are.  Thus we start with some initial choice and then, through the machine learning algorithm, train until we settle on encodings and decodings that minimize the loss function.

Let the putative encoding (decoding) for gate $g\in G$ be denoted $E^n[\mathbf{x}, g]$ ($D^n[\mathbf{x}, g]$) for the $n$th iterative step in the training for the octagon at $\mathbf{x}$.   We write $Y^n[\mathbf{x}, g] = (E^n[\mathbf{x},g], D^n[\mathbf{x},g])$. We are, then, admitting the possibility that the same gate might require different encodings and decodings in different regions of spacetime. One could imagine eventually adjusting the machine learning algorithm to be introduced to exploit any assumed homogeneity, but we will not explore this issue further. At the $n$th step we will have a particular encoding and decoding scheme
\begin{equation}
\{ (\mathbf{x}, Y^n[\mathbf{x}, g]):\forall~ \mathbf{x}\in X_\text{octagons},~ \forall ~ g\in G\}
\end{equation}
This specifies an encoding for every element of $G$ for every octagon. We will attempt to implement various quantum circuits using this encoding. Then, using the machine learning algorithm trained upon the resulting empirical information, we will iterate the encoding and decoding scheme, obtaining a new one to be used during the $(n+1)$th step.

We consider a fragment, $\mathcal{F}$,  made from the gates, $g$, in our UGS with locations assigned to the positions of some of the octagons in some octagon square tessellation.   
The fragment $\mathcal{F}$ is specified by
\[   \mathcal{F}=\{(\mathbf{x}, g): \forall \mathbf{x}\in X\subseteq X_\text{octagons} \}\]
where $X$ is the set of positions of octagons positions in the tessellation at which a gate is placed.  A circuit, $\mathcal{C}$, is 
a special case of a fragment in which every initial gate is of the type that ignores any open inputs into it. 

The attempted implementation of circuit $\mathcal{C}$ during the $n$th step is now given by
\begin{equation}
Y^n[\mathcal{C}]=\{ ( \mathbf{x},Y^n[\mathbf{x},g]): \forall \mathbf{x} \in X \}
\end{equation}
in the octagons. Actually, this is not entirely sufficient, since we must also specify what happens operationally in regions of spacetime outside the given octogons (tessels). We will return to this point at the end of this section. 

Each gate $g$ in $\mathcal{C}$ maps quantum information to classical outcomes with various probabilities. The full circuit, however, fixes this internal quantum information (by incuding state preparation and measurement ``gates", for example), and is thus characterized by a single probability $p_\mathcal{C}$ for a specific set of \emph{classical} outcomes to be observed. Quantum mechanics can be formulated \cite{OpTensor} as an assignment of a $p_\mathcal{C}$ to every possible circuit $\mathcal{C}$. 

In the next section, we will show that under certain assumptions, the converse is also true. That is, if each circuit within a \emph{tomographically complete circuit set} indeed occurs with its predicted $p_\mathcal{C}$, the underlying \emph{operational} map is indeed that specified by the relevant sequence of gates. 

In order to certify that the computation within the stuff is indeed quantum mechanical, we thus seek encoder/decoder pairs that perform this same assignment via the stuff. That is, the theoretical map, from circuit settings and gate labels $\mathcal{C}$ to outcome probabilities $p_\mathcal{C}$, is also the operational one, from the tesselation of $\mathcal{C}$ and a given $Y^n[\mathcal{C}]$ to the \emph{observed} outcome probability $p^n_\mathcal{C}$. 

We can then define one of several loss functions measuring how closely $Y^n[\mathcal{C}]$ indeed implements some $\mathcal{C}$,
\begin{equation}
\label{eq:LossSingle}
L[Y^n[\mathcal{C}], \mathcal{C}] = \mathrm{err}(p^n_\mathcal{C}, p_\mathcal{C}),
\end{equation}
where $\mathrm{err}(x, y)$ is some convenient positive function with a global minimum at $x=y$, for example $\mathrm{err} = |x^2 - y^2|$.

Now suppose we have fixed a set $\mathcal{S}$ of circuits that we wish to simultaneously implement, for example because they form a tomographically complete circuit set. A loss function over the full set can then be written as
\begin{equation}
\label{eq:LossSet}
\mathcal{L}[Y^n, \mathcal{S}] = \sum_{\mathcal{C} \in S} L[Y^n[\mathcal{C}], \mathcal{C}].
\end{equation}

Since the internal quantum information of the stuff might depend upon occurrences outside the spacetime region we have identified with our circuit, for this to work in practice
we must also specify what happens at the spacetime points, $\bar{X}$, that are not in the octagons labelled by $\mathbf{x}\in X$. This includes those in the squares and those that occur elsewhere.  

 One strategy out of potentially many is to choose a ``null" encoding $E_0$ at each ``external" spacetime point, and to simply ignore signals on the outcome wires there, so that we need not specify any decoding. The null encoding might also be iterated (``trained") in order to, for example, appropriately ``zero out" quantum information in the external regions; in that case it would be denoted $E^n_0$. Similar considerations apply to the squares.
 We do not need to concern ourselves with the encoding for spacetime points in the future of the circuit, $\mathcal{C}$, via the causal assumption that influences cannot travel backwards in time.  

\section{Bootstrap tomography}\label{sec:bootstrap}
Now let us discuss the construction of so-called ``tomographically complete circuit sets".

If the loss function, $\mathcal{L}[Y^n, \mathcal{S}]$, is minimized when summed over a big enough set of test circuits, we would like it to be the case that any circuit gives the correct probabilities (to within some small error).   In this section we state a theorem (proven in Appendix 1) that if the empirical probabilities, $p_\mathcal{C}^n$, are exactly equal to the ideal probabilities, $p_\mathcal{C}$ (so the loss function \eqref{eq:LossSet} is minimized), for a certain set of circuits, $\mathcal{S}_\text{tom}$, then this is true for all circuits.  The circuits in $\mathcal{S}_\text{tom}$ have the property that they are bounded in size.  We conjecture that a robust version of this theorem also holds - namely that the loss function \eqref{eq:LossSet} over $\mathcal{S}_\text{tom}$ bounds the loss function for all circuits.   

We will need the important notion of a \emph{bounded fragment (or circuit)}.  This is one that fits inside a box of some constant size, $\Delta L$ and $\Delta T$, where this box size does not increase in size as we increase $L$ and $T$.

It is shown in Appendix 1 that we can associate a vector, $\mathbf{r}^n_X[\mathcal{F}]$, with any fragment in region $X$.  This vector linearly relates the given fragment to a tomographically complete set of fragments for the given region.  In the case of Quantum Theory, the vector $\mathbf{r}_X[\mathcal{F}]$ is linearly related to the superoperator associated with the fragment. The vectors, $\mathbf{r}^n_X[\mathcal{F}]$, are used to calculate the probabilities for circuits.  

We can determine the vectors, $\mathbf{r}_X[\mathcal{F}]$, by doing tomography on a set of circuits $\mathcal{F}\cup\bar{\mathcal{F}}$ for different $\bar{\mathcal{F}}$.  We say we have \emph{fragment tomography boundedness} if we can do tomography on a bounded set of fragments (pertaining to $X$) by means of a bounded set of circuits.   

In Appendix 1 we define a \emph{composition tomograph}, $\pmb{\Lambda}_X^{X_1,X_2, \dots}$ which tells us how to combine $\mathbf{r}$ vectors pertaining to non-overlapping (though possibly adjacent) regions $X_1$, $X_2$, \dots to obtain the tomographic information pertaining to the region $X=X_1\cup X_2 \cup \dots$. 

We say we have \emph{composition tomography boundedness} if the composition tomograph for a composite region formed from any number of regions can be determined from the composition tomographs for composites that fit inside bounded boxes.  In other words, we can do the calculation for any circuit from calculations pertaining to smaller bounded parts of that circuit.

In Appendix 1 the following theorem is proven. 
\begin{quote}
\textbf{Theorem 1}.  If we have fragment tomography boundedness and composition tomography boundedness, then there exists a bounded set of circuits $ \mathcal{S}_\text{tom}$ such that if, at iteration $\tilde{n}$, we have
\[ p^{\tilde{n}}_\mathcal{C} = p_\mathcal{C}  ~~~\forall \mathcal{C}\in \mathcal{S}_\text{tom} \]
then $ p^{\tilde{n}}_\mathcal{C} = p_\mathcal{C}$ for any $\mathcal{C}\in \mathcal{S}_\text{circuits}$.
\end{quote}
$\mathcal{S}_\text{tom}$ is then called a ``tomographically complete circuit set". 
We conjecture (but do not prove) that a robust version of this theorem holds.
\begin{quote}
\textbf{Conjecture 1}.  If we have fragment tomography boundedness and composition tomography boundedness, then there exists a bounded set of circuits $ \mathcal{S}_\text{tom}$ such that if, at iteration $\tilde{n}$, we have
\[ \mathcal{L}[Y^{\tilde{n}}, \mathcal{S}_\text{tom}] \leq \epsilon \]
then
\[ \mathrm{err}(p^n_\mathcal{C}, p_\mathcal{C}) \leq \epsilon B N_\mathcal{C} ~~~~ \forall~ \mathcal{C} \]
where $B$ is a constant and $N_\mathcal{C}$ the number of gates in $\mathcal{C}$.
\end{quote}
The motivation for this conjecture is that the tomography process will fix the parameters in the gates to some error. If we enlarge the set, $\mathcal{S}_\text{tom}$, we can expect to get a better bound on the error since we then collect more information.   

The causaloid framework \cite{hardy2005probability,hardy2007towards,hardy2009quantum,markes2011entropy} is used to prove Theorem 1. This framework was originally developed as a framework for modelling indefinite causal structure in the context of Quantum Gravity.  
These theorems mean that we can rely on the quantum stuff to implement an arbitrary circuit as long as the measured probabilities for circuits in $\mathcal{S}_\text{tom}$ are close enough to the ideal probabilities calculated from Quantum Theory. This is good because it would not be practical to measure the probabilities for all $\mathcal{C}\in \mathcal{S}_\text{circuits}$ since the rank of this set grows very rapidly with $L$ and $T$.  

For the two tomographic boundedness properties to hold requires in each case that (i) that a mathematical prerequisite holds and then (ii) that the physics of the stuff accords.  The mathematical prerequisite is that the properties hold for ideal circuits constructed on the given lattice from the given universal gate set. To check this requires a mathematical calculation.  Since we have a universal gate set, it is immediately clear that this mathematical prerequisite holds for fragment tomography boundedness (as we can use the UGS to construct a tomographically complete set of fragments that are bounded).  We conjecture in Appendix 1 that the mathematical prerequisite holds for composition tomography boundedness for any UGS.  If the mathematical prerequisites hold, then we can consider whether the physical properties accord.  This will, most likely, be settled through working with the stuff.  However, we can always cook up situations in which the stuff fails to have the properties.  For example, the properties will fail if the stuff has hidden signalling between far separated locations.  For example, the bit of the stuff at $\mathbf{x}$ might send a radio frequency signal to the bit of stuff at some $\mathbf{x}'$ in the deep future outside any bounding box where this radio signal cannot tomographically probed by circuits that live in a bounding box.  

While it would be satisfying to prove Conjecture 1 mathematically, it is possibly more useful to test it empirically.  The conjecture does, in any case, necessarily involve assuming the boundedness properties which are, themselves, in need of empirical investigation. We test the conjecture empirically by determining the extent to which minimizing the loss function on various bounded sets of circuits allows us to reproduce the probabilities for sets of larger circuits. 

\section{Random Circuit Sampling}
In the preceding sections we have illustrated that minimization of the loss function \eqref{eq:LossSet} by an implementation $Y^n[\mathcal{C}]$, acting upon the stuff with respect to a tomographically complete set of quantum circuits $\mathcal{S}_\mathrm{tom}$, certifies that each individual gate in a UGS has also been correctly implemented by $Y^n[\mathcal{C}]$. The loss function \eqref{eq:LossSet} is inconvenient to use directly, however, because a) computing 
a probability for every circuit in $\mathcal{S}_\mathrm{tom}$ for every training iteration is likely to be expensive and b)
the differing contributions to the overall gradients by the (possibly many) terms in the sum over $\mathcal{S}_\mathrm{tom}$ are expected to confound gradient descent optimization. 

We will instead operate upon randomly constructed circuits, with each training iteration acting upon either a single example or perhaps a small minibatch of them. Random circuits can be constructed in various ways.  For example, we could start by randomly assigning gates from the UGS to a few positions, resulting in a fragment. Next, we can identify the locations of open wires, and close the circuit by assigning random preparation measurement gates to each. Alternatively, one could begin with the preparation gates, randomly add gates from the UGS capable of receiving inputs from those already present, and at some also-random point close the construction with a measurement gate, repeating the process from scratch if it has failed to deliver a circuit. Any such random constructions will be controlled by some parameters which dictate the average size of the circuits.  

Consider a set (minibatch), $\mathcal{S}_\text{rand}$, of circuits generated randomly by some such technique. We make the following random sampling assumption
\begin{quote}
\textbf{Assump: Random circuit sampling}.  There exists a bounded set of circuits $ \mathcal{S}_\text{rand}$ such that if, at iteration $\tilde{n}$, we have
\[ \mathcal{L}[Y^{\tilde{n}}, \mathcal{S}_\text{rand}] \leq \epsilon \]
then
\[ \mathrm{err}(p^n_\mathcal{C}, p_\mathcal{C}) \leq \epsilon D N_\mathcal{C} ~~~~ \forall~ \mathcal{C} \]
where $D$ is a constant and $N_\mathcal{C}$ the number of gates in $\mathcal{C}$.  
\end{quote}
For this assumption to hold good, we must have a machine learning algorithm that does not know what random set of circuits is going to be chosen from one iteration to the next. In other words, our procedure must indeed be ``random" in the sense that the algorithm never learns to predict its output in advance. 

The practical loss function is thus
\begin{equation}
\label{eq:LossRandom}
\mathcal{L}_\mathrm{rand}[Y^n, \mathcal{S}_\text{tom}] =\sum_{S_\mathrm{rand} \subset S_\mathrm{tom}} \sum_{\mathcal{C} \in S_\mathrm{rand}} L[Y^n[\mathcal{C}], \mathcal{C}],
\end{equation}
where $S_\mathrm{rand}$ is a randomly constructed subset of $S_\mathrm{tom}$. Since $S_\mathrm{rand}$ would canonically contain only one or a few elements, the interior sum is much simpler than that of \eqref{eq:LossSet}. In addition, by virtue of the random circuit sampling assumption, the exterior sum can be treated by individually optimizing each of its summands. In other words, repeatedly optimizing $L[Y^n[\mathcal{C}], \mathcal{C}]$ with respect to randomly constructed $\mathcal{C}$ from $S_\mathrm{tom}$ also optimizes \eqref{eq:LossRandom} and, by assumption, \eqref{eq:LossSet}.





\section{Machine Learning Algorithm}
\label{sec:MachineLearning}
\subsection{Neural network implementation of encoder/decoder}
We have repeatedly alluded to our intention to view the encoder/decoder pair as neural networks to be trained by a machine learning algorithm. In this section we will elaborate upon this process. Recall that in the preceding sections we have formulated the problem of translating arbitrary quantum programs into operations upon the stuff as \emph{bootstrap tomography}: the functional minimization of the loss function \eqref{eq:LossSet} (in practice \eqref{eq:LossRandom}) obtained by comparing the observed and predicted outcome distributions with respect to the encoder/decoder pair at training iteration $n$, $E^n[\mathbf{x}, g]$ and $D^n[\mathbf{x}, g]$. 
Here $g = g(t, x)$ is the gate label assigned to the point $\mathbf{x} = (t, x)$ by the tesselation of the given circuit.

The problem of automatically varying a function to optimize a loss function is the central concern of \emph{machine learning}. The subfield of \emph{deep learning} \cite{DeepLearning, GoodfellowDL} has become explosively popular in the past decade or so, as sufficient computational resources have become available to produce nearly magical - and highly profitable - results in the most impressive cases. Deep learning becomes especially useful relative to other approaches when the function to be optimized represents, or depends upon, a complicated probability distribution, as one might expect the encoder/decoder pair to.

A deep learning optimization, or ``training", begins by representing the target function as a linear composition of smooth \emph{activation functions} called a \emph{neural network}. Different \emph{network structures} are defined by different arrangements of activation functions, with ``deep" learning being somewhat vaguely defined by its focus upon network structures formed of ``many" successive layers. The activation functions themselves are parameterized by their \emph{weights} $\theta$ so that different functions decompose into a given network structure by different choices of weights. 

Every function theoretically has some neural network representation \cite{GoodfellowDL}. While this in itself is not especially impressive, the linearity and smoothness of a neural network's activation functions permit functional derivatives with respect to the full network to be expressed as sums of partial derivatives of the weights, essentially via the chain rule. The gradient of the full function with respect to some loss can then by efficiently descended by descending those of the individual activation functions, a process known as \emph{backpropagation} \cite{Backprop1, Backprop2, GoodfellowDL}. 

Precisely how backpropagation is best applied in a given situation is somewhat problem-dependent. We will save detailed consideration of this matter in future studies, when we implement bootstrap tomography on a small, classically simulated spin chain.

\begin{figure}
    \centering
    \includegraphics{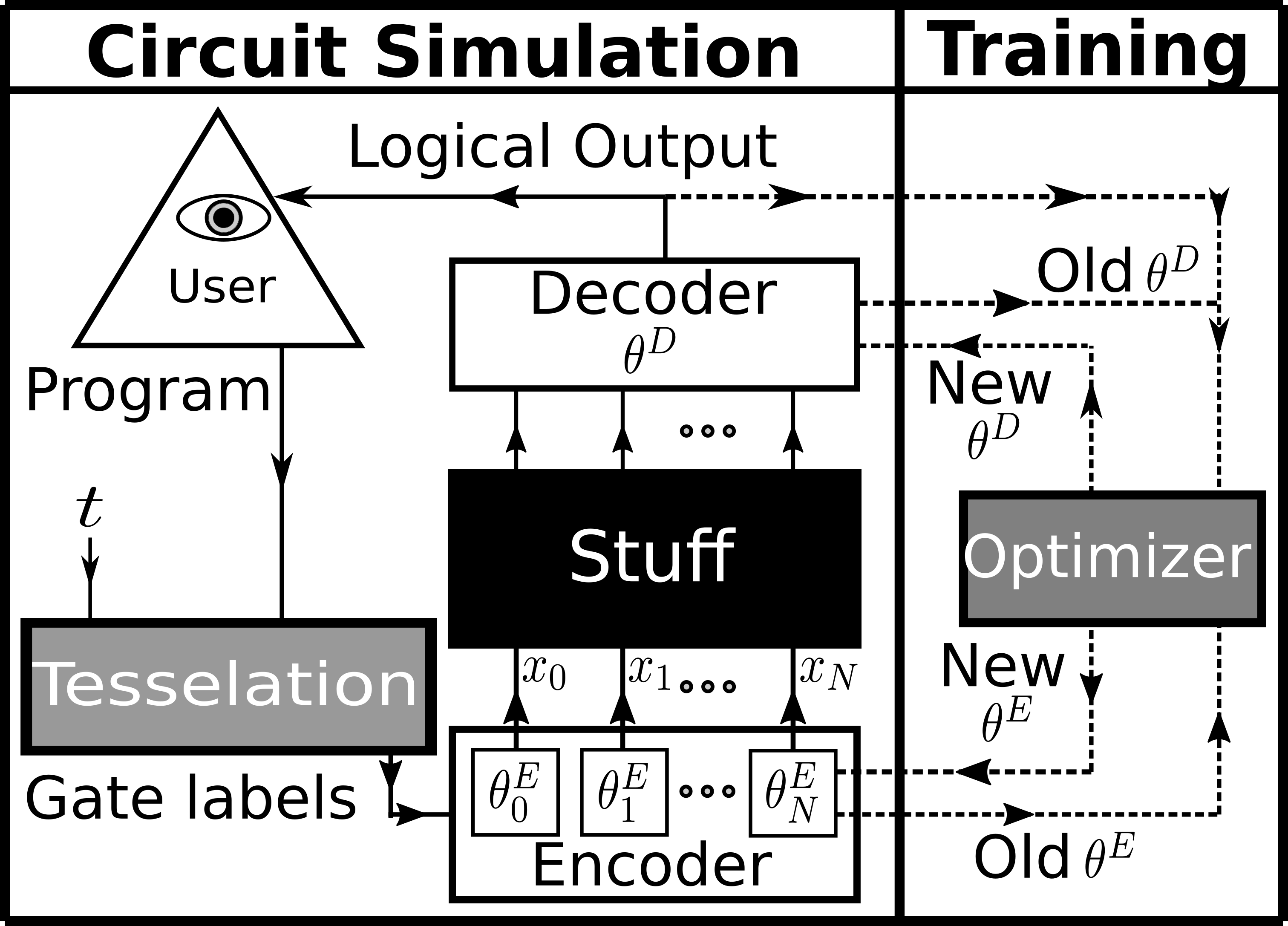}
    \caption{Depiction of the simulation algorithm. \textbf{Circuit Simulation:} The user decides on a program (circuit), which via the tesselation assigns a gate label to each spacetime event. Each \emph{spatial} point is assigned an encoder, mapping these gate labels to inputs to the stuff. Output from the stuff is then processed by the decoder into simulated logical output from the program. The decoder receives the gate label $g$ and the central spatial point of the tessel $x$ in addition to its depicted input. \textbf{Training:} by choosing circuits from a  tomographically complete circuit set, the comparison between predicted and actual output from each can be used as a loss function for the encoder and decoder, such that \emph{all} circuits are correctly implemented at optimum. This is achieved by representing the encoder and decoder as neural networks, and descending their weights towards this optimum, using randomly constructed circuits as input.}
    \label{fig:FlowChart}
\end{figure}

\begin{figure*}
\begin{minipage}{\linewidth}
\begin{algorithm}[H]
\begin{algorithmic}[1]
\Procedure{Optimize}{$\left\{\theta^E_{x}\right\}, \theta^D$}
\For{\emph{number of training iterations}}
\State $Circuit \gets$ a randomly generated circuit.
\State $LogicalOutput \gets $ a batch of $m$ calls to \sc SimulateCircuit($\left\{\theta^E_{x}\right\}, \theta^D, circuit$). \rm
\State $PredictedOutput \gets m$ predicted circuit outcomes.
\State $Loss \gets$ distance metric between $LogicalOutput$ and $PredictedOutput$ per \eqref{eq:LossSingle}.
\State Descend the gradients of $\left\{\theta^E_{x}\right\}$ and $\theta^D$ to minimize $Loss$, as dictated by the chosen optimization strategy.
\EndFor
\EndProcedure
\end{algorithmic}
\caption{Schematized optimization of the encoders and decoder.}\label{BootstrapAlgorithm}
\end{algorithm}
\end{minipage}
\end{figure*}

For now we will instead depict the problem at the more abstract level depicted in Figure \ref{fig:FlowChart}.
We will denote the neural network representation of the decoder as $\mathcal{D}[\theta^D; \mathbf{x}, g]$, so that
\begin{equation}
    D^n[\mathbf{x}, g] = \mathcal{D}[\theta^D; \mathbf{x}, g].
\end{equation}
The network representation $\mathcal{D}$ itself is fixed for all $n$. Training updates are instead implemented by varying the weights $\theta^D$ between training iterations - they are otherwise fixed.

The encoder $E^n[\mathbf{x}, g]$ controls the internal quantum information of the stuff, and thus unlike the decoder must interact with it in real time. It nevertheless needs to share information within tessels in order to track that quantum information's flow. This point will be elaborated upon in the upcoming Subsection B. For now we briefly note that we handle this problem by representing $E^n[\mathbf{x}, g]$ with a ``fleet" ``recurrent" neural networks, one at each spatial point $x$. We denote the encoder weights at some $x$ as $\theta^E_x$, and the full set of weights over all spatial points as $\left\{\theta^E_x\right\}$. Thus
\begin{equation}
    E^n[\mathbf{x}, g] = \mathcal{E}[\left\{\theta^E_x\right\}, \mathcal{M}; \mathbf{x}, g].
\end{equation}
The network representation $\mathcal{E}$ is again fixed for all training iterations $n$, with variation between iterations instead implemented by manipulations of the weights $\left\{\theta^E_x \right\}$. The output of the network additionally depends upon a \emph{memory vector} $\mathcal{M}$. This vector is passed between different $\theta^E_x$ within a tessel, allowing encoders following different constant-$x$ ``worldlines" to communicate. We will elaborate upon this point in Subsection B.

Let us now follow the logic of Figure \ref{fig:FlowChart} in words. The user selects a program, which is mapped by the tesselation into a field of gate labels $g(t, x)$. Each gate label at $x$ is passed along with the appropriate memory vector $\mathcal{M}$ to the encoder at that same $x$, $\mathcal{E}[\theta^E_x, \mathcal{M}; \mathbf{x}, g]$. This yields a raw input signal, to be sent to the setting wire at $x$. Once an entire tessel has been implemented, the corresponding ``raw" outcome settings from the stuff are passed into the decoder, which maps them into ``logical" output.

If the weights $\theta^D$ and $\left\{\theta^E_x\right\}$ are optimal, as indicated by the loss function \eqref{eq:LossRandom} reaching its global minimum, the logical output may be interpreted as the correct result of the program. Otherwise, the gradients of the weights in the direction of decreasing \eqref{eq:LossRandom} can be calculated, as by the ``optimizer" in Figure \ref{fig:FlowChart}, and descended along to obtain new, better optimized weights.  In machine learning parlance, this process is called ``training" the networks. The next iteration operates upon a new randomly constructed circuit, and the process is repeated until a desired convergence threshold is reached. Illustrative pseudocode is provided as Algorithm \ref{BootstrapAlgorithm}.

\begin{figure*}
    \includegraphics{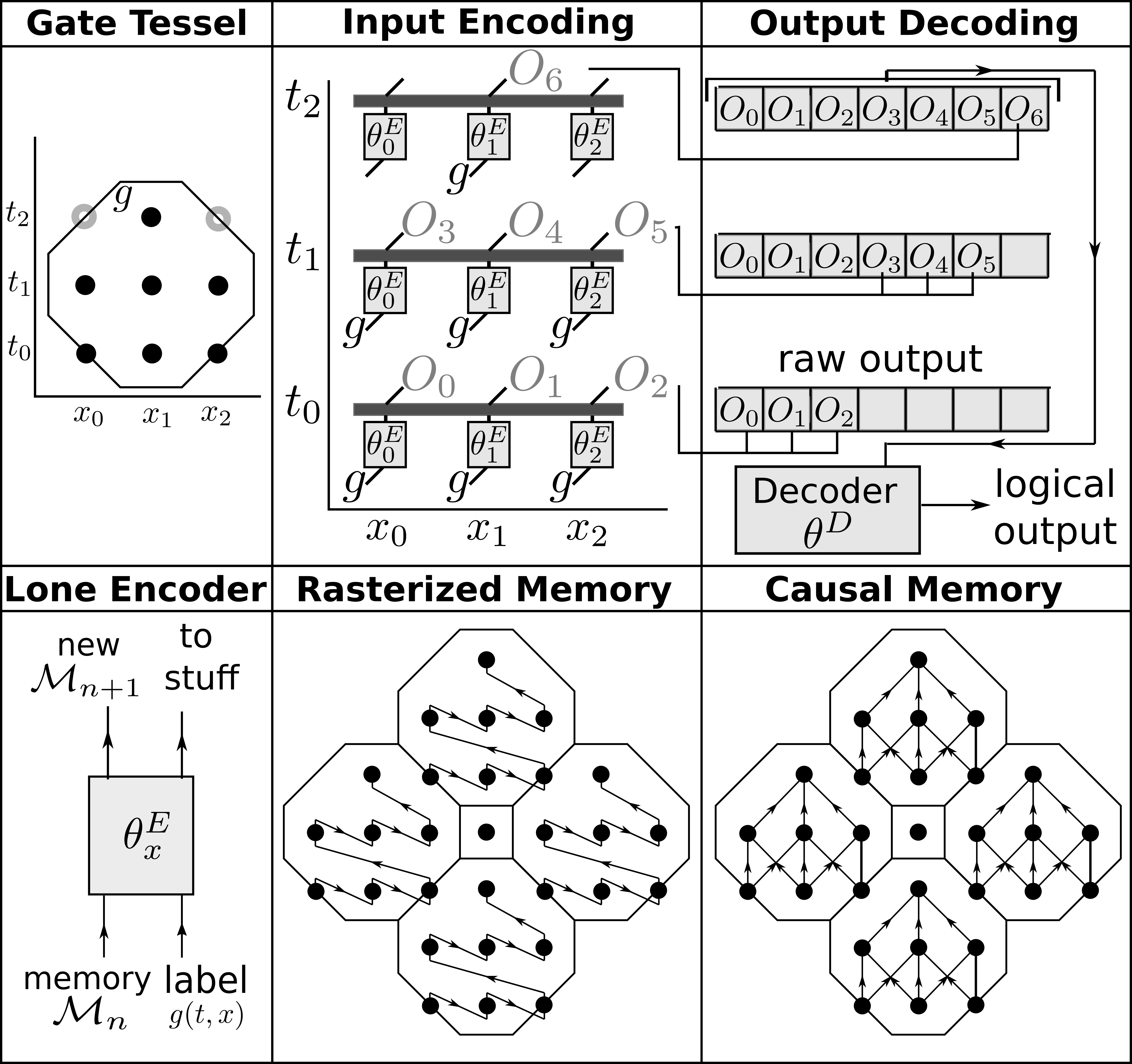}
    \caption{\textbf{Top Left:} each point $(t, x)$ in spacetime is assigned a gate label $g$ by the tesselation. $g$ is constant within each tessel, and takes a uniform ``null" value outside of a tessel. \textbf{Bottom Left:} each spatial point $x$ is assigned a recurrent neural network (RNN) ``Encoder", mapping $g$ and a ``memory" vector $\mathcal{M}$ to the input to the stuff, along with a new $\mathcal{M}_{n+1}$ (note this $n$ paramaterizes subsequent RNN calls, not training iterations). The map is governed by ``weights" $\theta^E_x$, local to $x$ and held fixed except during training. \textbf{Top Middle, Top Right:} the stuff advances through time, receiving encoded input dictated by the gate labels. Its raw outputs $\mathcal{O}(t, x)$ at each point in each tessel are collected into a vector, and then fed along with the gate label $g$ to the decoder. The decoder, another neural network with weights $\theta^D$, emits the simulated ``logical output" of the gate. \textbf{Bottom Middle, Bottom Right:} two strategies 
    to allow the encoder RNNs to collaborate over a region of spacetime. Rasterized memory (Bottom Middle) involves passing $\mathcal{M}_n$ in left-right order throughout a tessel. Causal memory (Bottom Right) involves passing it forward within each encoder's future light cone, achieving, per the locality assumption, the same end in a shorter timescale.}
    \label{fig:MemoryFig}
\end{figure*}

\subsection{RNN fleet implementation of encoder}
\label{FleetSubsection}
As discussed previously, the encoder $E^n[\mathbf{x}, g]$ and decoder $D^n[\mathbf{x}, g]$ have differing relationships with the real time behaviour of the stuff, which suggests a particular network implementation for the encoder that we call an ``RNN fleet". We will elaborate upon this point here. 

Figure \ref{fig:MemoryFig} depicts several aspects of the behaviour of the encoder/decoder algorithm in spacetime. In the top left panel, we see a single tessel, assigned to a gate labelled $g$. The top middle panel depicts this same tessel, implemented as operations upon the encoders and thus upon the stuff. As the stuff proceeds through time, the gate label $g$ is fed to the encoder at each $(t, x)$ in the tessel. As we see in the top right panel, the output from the tessel is collected into a vector. Once all of it has been collected, it is sent to the decoder, which produces the logical output of the gate.

Since nothing in this procedure depends upon the logical output directly, the precise time at which the decoder $D^n[\mathbf{x}, g]$ operates is not especially important, within reason. The encoder $E^n[\mathbf{x}, g]$, on the other hand, controls the internal quantum information of the stuff. The order in which its inputs are processed therefore critical. In addition, it must be synchronized within a tessel.

The problem of processing ``time series" with a specific temporal ordering occurs repeatedly in machine learning. The prototypical example is machine translation, since correct translations depend upon prior context. A genre of network structures known as ``recurrent neural networks", or RNNs \cite{Backprop1, LSTM, RNNReview, GoodfellowDL}, are adapted to remember such context. In addition to their weights, which are fixed except during training, such networks maintain a ``memory" vector $\mathcal{M}$. The output of the RNN depends on the memory as well as upon the weights and the input. But unlike the weights, the memory is modified between successive function calls, allowing it to represent contextual information: \emph{which} contextual information to record is in turn determined by the weights.

We thus implement the encoder $E^n[\mathbf{x}, g]$   at each separate spatial point $x$ as an RNN = $\mathcal{E}[\theta^E_x, \mathcal{M}; \mathbf{x}, g]$. as depicted in the bottom left panel of Figure \ref{fig:MemoryFig}. Each $\mathcal{E}[\theta^E_x, \mathcal{M}; \mathbf{x}, g]$ follows a particular constant-$x$ worldline through the tesselation, partly motivating the term ``RNN fleet". 

Additional motivation for the term comes from the need of the various RNNs to be synchronized within a tessel. Thus, the memory vector $\mathcal{M}_n$ is not simply passed forward along a worldline, but is instead shared within each tessel between networks of different weights. Consequently, we need to synchronize the processing of \emph{several} time series at different spatial points.

The bottom middle and bottom right panels depict two strategies for synchronizing within tessels. We call the first and simplest strategy, in the bottom middle panel of \ref{fig:MemoryFig}, \emph{rasterized memory}. In this paradigm, encoders $\mathcal{E}[\theta^E_x, \mathcal{M}; \mathbf{x}, g]$ are called sequentially at each timestep from left to right. Starting from a fixed null value, memory is thus passed within the zig-zagging ``spacelike" lines depicted in the bottom middle panel of Figure \ref{fig:MemoryFig}, which resemble the ``rasterized" beam path of a cathode ray tube television. Circuit simulation using the rasterized strategy is illustrated by the pseudocode of Algorithm \ref{SimulateCircuit}.

The illusion of motion created by such televisions relies upon the time required for the beam to traverse the screen being much shorter than the processing time of the eye. The ability of the rasterized memory strategy to effectively synchronize encoders within a tessel correspondingly depends upon the processing time of the encoders being much shorter than that between simulation timesteps.

The \emph{causal memory} strategy depicted in the bottom right of Figure \ref{fig:MemoryFig} relaxes this assumption. It is based on the assumption that the error incurred by sending encoder input out of sequence falls off with the spacetime distance between the disordered events. 

Instead of passing memory sideways between every event in the tessel, we thus pass it forward in spacetime within a fixed ``light cone" of predetermined width $s$, unless doing so would cross a tessel. Given locality, this should synchronize just as well as the rasterized strategy, without putting impositions upon the relative timescales of the encoder and the simulations. Circuit simulation using the causal strategy is illustrated by the pseudocode of Algorithm \ref{SimulateCircuitCausal}.

\begin{figure*}
\begin{minipage}{\linewidth}
\begin{algorithm}[H]
\begin{algorithmic}[1]
\Procedure{SimulateCircuitRasterizedMemory}{$\left\{\theta^E_x\right\}, \theta^D, circuit$}
\State $g(t, x) \gets$ the gate labels assigned to each spacetime point by the tesselation of $circuit$.
\State $n_T \gets$ a separate point counter for each tessel $T$ with fixed $g$.
\State $\mathcal{M}^T_{n_T} \gets$ a separate memory vector for each $T$.
\For{$t$ in the tesselation}
	\For{$x$ in the tesselation}
	    \State Determine which tessel $T$ the point $(t, x)$ inhabits.
	    \State Retrieve this $g$, $n_T$, and $\mathcal{M}^T_{n_T}$.
	    \State $InputSignal, M^T_{n_T+1} \gets$ output mapped from $g$ and $\mathcal{M}^T_{n_T}$ by encoder at $x$ with weights $\theta^E_x$.
		\State $RawOutput[n_T] \gets $ the corresponding output from stuff at $x$.
		\State $n_T \gets n_T + 1$.
	\EndFor
	\If{\rm{this is the last point in a tessel}}
		\State $LogicalOutput[T] \gets$ output mapped from $RawOutput$, $g$, and $x$ by decoder with weights $\theta^D$.
	\EndIf
\EndFor
\State \Return $LogicalOutput$
\EndProcedure
\end{algorithmic}
\caption{Circuit simulation using \emph{rasterized} memory, passed sequentially between each point in each tessel.}\label{SimulateCircuit}
\end{algorithm}
\end{minipage}
\end{figure*}

\begin{figure*}
\begin{minipage}{\linewidth}
\begin{algorithm}[H]
\begin{algorithmic}[1]
\Procedure{SimulateCircuitCausalMemory}{$\left\{\theta^E_x\right\}, \theta^D, circuit$, light cone width $s$.}
\State $g(t, x) \gets$ the gate labels assigned to each spacetime point by the tesselation of $circuit$.
\State $n_T \gets$ a separate point counter for each tessel $T$ with fixed $g$.
\State $\mathcal{M}_{t=0, x} \gets$ a separate memory vector for each spatial point $x$.
\For{$t$ in the tesselation}
	\For{$x$ in the tesselation}
	    \State Retrieve $g(t, x)$ and $\mathcal{M}_{t-s, x-s:x+s}$, the memories between $x-s$ and $x+s$.
	    \State $\tilde{\mathcal{M}}_{t, x} \gets$ pointwise multiplication of  $\mathcal{M}_{t-s, x-s:x+s}$ within the same tessel as $(t, x)$.
	    \State $InputSignal, M_{t, x} \gets$ output mapped from $g$ and $\tilde{\mathcal{M}}_{t, x}$ by encoder at $x$ with weights $\theta^E_x$.
		\State $RawOutput[n_T] \gets $ the corresponding output from stuff at $x$.
		\State $n_T \gets n_T + 1$.
	\EndFor
	\If{\rm{this is the last point in a tessel}}
		\State $LogicalOutput[T] \gets$ output mapped from $RawOutput$, $g$, and $x$ by decoder with weights $\theta^D$.
	\EndIf
\EndFor
\State \Return $LogicalOutput$
\EndProcedure
\end{algorithmic}
\caption{Circuit simulation using \emph{causal} memory, passed within forward lightcones, confined by tessels.}
\label{SimulateCircuitCausal}
\end{algorithm}
\end{minipage}
\end{figure*}

\section{Acknowledgements}
A. G. M. Lewis is supported by the Tensor Network Initiative at Perimeter Institute. Research at Perimeter Institute is supported by the Government of Canada through the Department of Innovation, Science and Economic Development Canada and by the Province of Ontario through the Ministry of Research, Innovation and Science.

\section*{Appendix 1}

In this appendix we provide definitions of fragment tomography boundedness and composition tomography boundedness and we prove Theorem 1.  The idea is to consider tomography on fragments.  We consider two types of tomography. First, we have \emph{fragment tomography}, whereby we obtain a mathematical object ($\mathbf{r}_X[\mathcal{F}]$ below) associated with the fragment.   Second, we consider \emph{composition tomography} whereby we obtain the rule for composing these  mathematical objects for composite regions such as $X_1\cup X_2\cup X_3$.  If the tomography boundedness properties hold then we only need to consider fragment tomography for fragments up to a certain size and composition tomography for composites up to a certain size.  To do tomography on these fragments and composite fragments are completed into circuits.  Hence it follows that we only need to consider circuits need be no bigger than a certain size.  This provides our set, $\mathcal{S}_\text{tom}\subset \mathcal{S}_\text{circuits}$ of circuits.   If we obtain probabilities $p^n_\mathcal{C}=p_\mathcal{C}$ for $\mathcal{C}\in \mathcal{S}_\text{tom}$ (to within some bounded error) then it follows that $p^n_\mathcal{C}=p_\mathcal{C}$ (to within some bounded error) for all circuits.

For a circuit, $\mathcal{C}=\mathcal{F}\cup\bar{F}$ we can always write the probability as
\begin{equation}\label{rdotp}
p^n_{\mathcal{F}\cup\bar{\mathcal{F}}} = \mathbf{r}^n_{X}[\mathcal{F}]\cdot \mathbf{p}^n_{X}[\bar{\mathcal{F}}]
\end{equation}
where we define the ordered set
\begin{equation}
\mathbf{p}^n[\bar{\mathcal{F}}] = ( p^n_{\mathcal{F}^k\cup\bar{\mathcal{F}}}: \text{for}~ k\in\Omega_{Y})
\end{equation}
where $k\in\Omega_{X}$ labels the elements, $\mathcal{F}^k$, of some tomographic set, $\mathcal{T}^\text{tom}_{X}\subseteq \mathcal{T}_{X}$ of minimal possible rank.  Here $\mathcal{T}_{X}$ is the set of all possible fragments at $X$.  In general, the choice of tomographic set, $\mathcal{T}^\text{tom}_{X}$, is not unique.  We can always write \eqref{rdotp} because, in the worst case, we can choose $\mathcal{T}^\text{tom}_{X}= \mathcal{T}_{X}$ and then the vector $\mathbf{r}^n_{Y}[\mathcal{F}]$ is just a list of $0$'s except with a $1$ at position $k$.  In general, however, there will be some linear relationships between these probabilities so that we can use a proper subset, $\mathcal{T}^\text{tom}_{X}\subset \mathcal{T}_{X}$.   We can think of $\mathbf{p}^n_{X}[\bar{\mathcal{F}}]$ as the \emph{generalized state} prepared by $\bar{\mathcal{F}}$ for region $X$.  And we can think of $\mathbf{r}^n_{X}[\mathcal{F}]$ as the \emph{generalized effect} associated with fragment $\mathcal{F}$ performed in region $X$.   The choice of tomographic set, $\mathcal{T}^\text{tom}_{X}$, must be good for calculating the probability for any circuit in any region $X\cup\bar{X}$ (that is for any $\bar{X}$ associated with a fragment $\bar{\mathcal{F}}$).   

Using simple linear algebra \cite{mana2003can} we can obtain the set $\{\mathbf{r}_X^n[\mathcal{F}]: \forall ~\mathcal{F}\in\mathcal{T}_X \}$ if we are given enough empirical information in the form of $p^n_\mathcal{C}$ for $\mathcal{C}\in S_{X}^\text{tom}$.  The set, $S_X^\text{tom}$, has to be big enough to make this possible - namely it has to generate $\Omega_{X}$ linearly independent $\mathbf{p}^n_{X}[\bar{\mathcal{F}}]$ vectors.  In this case we will say $S_X^\text{tom}$ is \emph{tomographically complete} for $X$.   We can be sure this is true simply by choosing $\mathcal{S}_X^\text{tom}$ to be the set of  all circuits in $\mathcal{S}_\text{all circuits}$ having elements at positions in $X$.  This is not very useful however as this set grows very rapidly with $L$ and $T$.

To obtain a more useful notion we define a \emph{bounded set} of circuits or fragments as one for which each element fits inside a box with bounded spatial and temporal dimensions which do not scale with $L$ and $T$.  Consider the following property.
\begin{quote}
\textbf{Fragment tomography boundedness}.  We say we have fragment tomography boundedness if,  for any bounded set of fragments,  there exists a bounded and tomographically complete set of circuits.
\end{quote}
We can motivate the assumption that this property holds by finiteness and locality.  First, note that the operationally accessed part of the Hilbert space associated with the inputs and outputs for any $\mathcal{F}\in\mathcal{T}$ should be finite so only require a finite number of circuits for tomography.  Furthermore, by locality we should be able to do tomography on this Hilbert space by means of circuits that are not too much bigger than the fragments.  

Consider a composite region, $X_1\cup X_2$ (where $X_1$ and $X_2$ are disjoint).   Then, for the circuit $\mathcal{C}=\mathcal{F}_1\cup\mathcal{F}_2\cup\bar{\mathcal{F}}$, we can write the probability as
\begin{equation}
p^n_{\mathcal{F}_1\cup\mathcal{F}_2\cup\bar{\mathcal{F}}} =
\mathbf{r}^n_{X_1\cup X_2}[\mathcal{F}_1\cup\mathcal{F}_2]\cdot \mathbf{p}^n_{X_1\cup X_2}[\bar{\mathcal{F}}]
\end{equation}
where
\begin{equation}
\mathbf{p}^n_{X_1\cup X_2}[\bar{\mathcal{F}}] = ( p^n_{\mathcal{F}^{k_1}\cup\mathcal{F}^{k_2}\cup\bar{\mathcal{F}}}: \text{for} ~ k\in\Omega_{X_1\cup X_2})
\end{equation}
It can be shown \cite{hardy2005probability} that for the composite region $X_1\cup X_2$ we can choose a tomographic set of fragments of the form $\mathcal{F}^{k_1}\cup\mathcal{F}^{k_2}$ where
\begin{equation}
\Omega_{X_1\cup X_2} \subseteq \Omega_{X_1} \times \Omega_{X_2}
\end{equation}
and, correspondingly, that
\begin{equation}
\mathbf{r}^n_{X_1\cup X_2}[\mathcal{F}_1\cup\mathcal{F}_2]  
=\pmb{\Lambda}_{X_1\cup X_2}^{X_1, X_2}
 (\mathbf{r}^n_{X_1}[\mathcal{F}_1] \otimes  \mathbf{r}^n_{X_2}[\mathcal{F}_2])
\end{equation}
where $\pmb{\Lambda}_{X_1\cup X_2}^{X_1,X_2}$ (which we called a composition tomogram in Sec.\ \ref{sec:bootstrap}) linearly projects vectors, $\mathbf{r}^n_{X_1}[\mathcal{F}] \otimes  \mathbf{r}^n_{X_2}[\mathcal{F}]$ living in a space of dimension $|\Omega_{X_1} \times \Omega_{X_2}|$ down to the vectors $\mathbf{r}^n_{X_1\cup X_2}[\mathcal{F}_1\cup\mathcal{F}_2]$ living in a space of dimension $|\Omega_{X_1\cup X_2}|$.
If we do fragment tomography to obtain $\mathbf{r}^n_{X_1}[\mathcal{F}_1]$,  $\mathbf{r}^n_{X_2}[\mathcal{F}_2]$, and $\mathbf{r}^n_{X_1\cup X_2}[\mathcal{F}_1\cup\mathcal{F}_2]$ then we can find an appropriate $\Omega_{X_1\cup X_2}$ set and solve for $\pmb{\Lambda}_{X_1\cup X_2}^{X_1,X_2}$.  This last step completes the composition tomography for $X_1\cup X_2$.

Composition tomography extends in the obvious way to more than two regions.  Thus, we can obtain $\pmb{\Lambda}_{X_1\cup X_2 \cup X_3}^{X_1,X_2,X_3}$, $\pmb{\Lambda}_{X_1\cup X_2\cup X_3 \cup X_4}^{X_1,X_2, X_3, X_4}$, and so on.   It turns out that these $\pmb{\Lambda}$'s are related by mathematical identities if the $\Omega$ sets associated with the $X$'s have certain relationships with one another (see Sec.\ 23 of \cite{hardy2005probability}).   Consider the following property
\begin{quote}
\textbf{Composition tomography boundedness}.  We will say we have composition tomography boundedness for circuits formed on a given lattice using a given gate set, $G$, if we can obtain $\pmb{\Lambda}_X^{X_1,X_2, \dots}$ for any $X$ from mathematical identities concerning only $\pmb{\Lambda}$'s pertaining to boxes bounded by some constant size $(\Delta L_\text{bounded}, \Delta T_\text{bounded})$ (which do not increase with $L$ and $T$).
\end{quote}
Here we say that $\pmb{\Lambda}_{X_1\cup X_2\cup \dots \cup X_M}^{X_1,X_2, \dots, X_M}$ pertains to a box bounded by $(\Delta L_\text{bounded},\Delta T_\text{bounded})$ if $X_1\cup X_2\cup \dots \cup X_M$ fits inside such a box.  From the results in  Sec.\ 29 of \cite{hardy2005probability} it is clear this conjecture is true for a specific gate set, $G$, consisting of gates for which the associated operators span the full space of operators acting on the Hilbert space associated with two qubits.  In this case, $|G|=256$. The gate set illustrated in the above table has only $|G|=14$.   We conjecture that circuits formed on the above lattice with this gate set also have the property of composition tomography boundedness.  More generally, we conjecture that circuits formed with respect to any lattice with any universal gate set have this property.

We can calculate the probability for any circuit, $\mathcal{C}$ using only $\mathbf{r}$ vectors in the following way.  First we consider
\begin{equation}  \label{pequalsrpC}
p^n_{\mathcal{F}\cup\bar{\mathcal{F}}}= \mathbf{r}^n_X[\mathcal{F}]\cdot \mathbf{p}^n_X[\bar{\mathcal{F}}]
\end{equation}
where $X$ is the region for fragment $\mathcal{F}$, and $\bar{\mathcal{F}}$ is another fragment.   Also consider
\begin{equation} \label{pequalsrpCequalsone}
\text{prob}(\bar{\mathcal{F}}|\mathcal{T}_\text{choice}) = \sum_l p^n_{\mathcal{F}[l]\cup\bar{\mathcal{F}}}
= \sum_l \mathbf{r}^n_X[\mathcal{F}[l]]\cdot \mathbf{p}^n_X[\bar{\mathcal{F}}]
\end{equation}
where $\mathcal{F}=\mathcal{F}[1]$ and $\mathcal{T}_\text{choice}=\{\mathcal{F}[l]: \forall l\}$ are a set of fragments for $X$ with mutually exclusive and exhaustive set of outcomes so that there probabilities must add to one.  We denote this set with the subscript \lq\lq choice" because it represents the choice we make in $X$ (a different choice would correspond to a different mutually exclusive and exhaustive set).  The vector $\mathbf{p}_X[\bar{\mathcal{F}}]$ depends on $\bar{\mathcal{F}}$ and is independent of $\mathcal{F}[l]$ in region $X$.   By Bayes rule
\begin{equation}
\text{prob}^n(\mathcal{F}|\bar{\mathcal{F}},\mathcal{T}_\text{choice})
= \frac{  \mathbf{r}^n_X[\mathcal{F}]\cdot \mathbf{p}^n_X[\bar{\mathcal{F}}]}
{\sum_l \mathbf{r}^n_X[\mathcal{F}[l]]\cdot \mathbf{p}^n_X[\bar{\mathcal{F}}]}
\end{equation}
Hence, iff
\begin{equation}\label{propcond}
\mathbf{r}^n_X[\mathcal{F}]~~~ \text{is proportional to}~~~ \sum_l \mathbf{r}^n_X[\mathcal{F}[l]]
\end{equation}
then $\text{prob}^n(\mathcal{F}|\bar{\mathcal{F}},\mathcal{T}_\text{choice})$ is independent of $\bar{\mathcal{F}}$. We denote this probability by $p^n_{\mathcal{F}|\mathcal{T}_\text{choice}}$ and it is given by
\begin{equation}
\mathbf{r}^n_X[\mathcal{F}] = p^n_{\mathcal{F}|\mathcal{T}_\text{choice}}  \sum_l \mathbf{r}^n_X[\mathcal{F}[l]]
\end{equation}
We can apply this approach to the case where $\mathcal{F}$ and $\bar{\mathcal{F}}$ are circuits.   If we have ideal circuits then it is true in Quantum Theory (and circuit theories in general \cite{hardy2011reformulating}) that
\begin{equation}\label{pCpCbar}
p_{\mathcal{C}\cup\bar{\mathcal{C}}} = p_\mathcal{C} p_{\bar{\mathcal{C}}}
\end{equation}
From this property it can be proven that the proportionality condition in \eqref{propcond} must hold and hence the probability for a circuit is given by
\begin{equation}
\mathbf{r}_X[\mathcal{C}] = p_\mathcal{C} \sum_l \mathbf{r}_X[\mathcal{C}[l]]
\end{equation}
Here we have dropped the possible dependence of $p_\mathcal{C}$ on the choice of mutually exclusive circuit set  (it is possible to prove that there is no such dependence).  
For non-ideal circuits, however, the proportionality condition of \eqref{propcond} may fail even when $\mathcal{F}$ is a circuit because we have not fully trained it to behave like a circuit.    Nevertheless, if the proportionality condition holds approximately, then we can bound the probability \cite{markes2011entropy}.

We now prove the following theorem.
\begin{quote}
\textbf{Theorem 1}.  If we have fragment tomography boundedness and composition tomography boundedness, then there exists a bounded set of circuits $ \mathcal{S}_\text{tom}$ such that if, at some iteration $n=\tilde{n}$, we have
\[ p^{\tilde{n}}_\mathcal{C} = p_\mathcal{C}  ~~~\forall \mathcal{C}\in \mathcal{S}_\text{tom} \]
then $ p^{\tilde{n}}_\mathcal{C} = p_\mathcal{C}$ for any $\mathcal{C}\in \mathcal{S}_\text{circuits}$.
\end{quote}
Recall that $p_\mathcal{C}$ are the probabilities for idealised circuits when the gates $g$ are actually those in our UGS, $G$.  We can obtain these probabilities by calculation using Quantum Theory.  
To calculate $ p^{\tilde{n}}_\mathcal{C}$ for an arbitrary circuit we need to be able to calculate the $\mathbf{r}^n_X$ vectors for the region associated with $\mathcal{C}$.  We can do this in the following way.  If we have $\pmb{\Lambda}_X^{X_1,X_2, \dots}$ where $X_1$, $X_2$, \dots live inside bounded boxes and we can also calculate $\mathbf{r}_{X_1}$, $\mathbf{r}_2$, \dots for these same bounded boxes and then we can calculate $\mathbf{r}_X$ vectors for region $X$, and from these we can calculate $p_\mathcal{C}$.  We can obtain these $\mathbf{r}_{X_i}$ vectors from probabilities for bounded circuits by fragment tomography boundedness.  If we have composition tomography boundedness then we can obtain $\pmb{\Lambda}_X^{X_1,X_2, \dots}$ from $\pmb{\Lambda}$'s that pertain to bounded boxes.  Furthermore, we can determine these $\pmb{\Lambda}$'s that live in bounded boxes by constructing circuits that live in bigger, but still bounded, boxes by fragment tomographic locality.   Thus, all the circuits we need to do tomography live in bounded boxes.   If we obtain the probabilities $p^{\tilde{n}}_\mathcal{C} = p_\mathcal{C}$ for this set of circuits then, by the above argument, we will obtain $\mathbf{r}_X^n[\mathcal{F}]=\mathbf{r}_X[\mathcal{F}]$ for any $X$ and $\mathcal{F}$.  We can calculate the probability for any circuit using these $\mathbf{r}$ vectors and hence this proves the theorem.

\bibliography{plasticinePRLformat}
\end{document}